# Diagonalization-free implementation of spin relaxation theory for large spin systems

Ilya Kuprov

*Oxford e-Research Centre, University of Oxford,*
*7 Keble Road, Oxford OX1 3QG, UK.*

Fax: +44 1865 610612

Email: ilya.kuprov@oerc.ox.ac.uk

**ABSTRACT**

The Liouville space spin relaxation theory equations are reformulated in such a way as to avoid the computationally expensive Hamiltonian diagonalization step, replacing it by numerical evaluation of the integrals in the generalized cumulant expansion. The resulting algorithm is particularly useful in the cases where the static part of the Hamiltonian is dominated by interactions other than Zeeman (*e.g.* in quadrupolar resonance, low-field EPR and Spin Chemistry). When used together with state space restriction tools, the algorithm reported is capable of computing full relaxation superoperators for NMR systems with more than 15 spins.

**Keywords**

NMR, EPR, relaxation, simulation, spin dynamics

# I. Introduction

The semi-classical description of relaxation in spin systems, in which a one-way coupling to the lattice is introduced using perturbation theory [1-5], is very popular in magnetic resonance because it provides readily interpretable data on structure and dynamics of many interesting systems, such as biomolecules and polymers [6]. The most convenient perturbative relaxation theory [2,4] is well adapted for liquid-state NMR / ESR spectroscopy and uses second-order time-dependent perturbation treatment. In recent decades, it has been refined, particularly in the spectral density part [7-8], into a very powerful tool for the investigation of molecular structure and dynamics [6]. In its present form, the second-order theory is mostly credited to Redfield [4], who usually points at an earlier paper by Wangsness and Bloch [5] – hence the BRW abbreviation. Ultimately, all perturbative relaxation theories with classical lattices are special cases of a very powerful formalism of generalized cumulant expansions derived by Kubo and Freed [9-10].

As interesting spin systems grew larger over the years, one specific computational bottleneck has emerged in the otherwise excellent BRW theory – the requirement that the basis operators for the expansion of the dynamic part of the Hamiltonian be the eigenoperators of the static Hamiltonian commutation superoperator [1-2,4]:

$$\hat{H}_1(t) = \sum_k f_k(t)\hat{V}_k \qquad [\hat{H}_0, \hat{V}_k] = \hat{\hat{H}}_0 \hat{V}_k = \omega_k \hat{V}_k \qquad (1)$$

(*e.g.* Equation 2.2 in the original formulation [4] and Equations 12-14 in the recent review by Goldman [2]). In the case where the static Hamiltonian is dominated by Zeeman interactions (high-field NMR and ESR spectroscopy), these are just irreducible spherical tensors [11]:

$$[\hat{L}_Z, \hat{T}_{lm}] = m\hat{T}_{lm}, \qquad (2)$$

but in most other cases these operators are unknown, and the Hamiltonian superoperator must be diagonalized to obtain them. In situations where matrix dimension exceeds $n = 10^4$, diagonalization is not possible because of the $O(n^3)$ multiplications required and because the eigenvectors of sparse matrices are in most cases dense and overflow the computer memory.

In this communication, a simple solution to this problem is suggested, using the fact that evaluation of the exponential propagator $\exp(-i\hat{\hat{H}}_0 \Delta\tau)$ for $\Delta\tau \sim \|\hat{\hat{H}}_0\|^{-1}$ is cheaper than $\hat{\hat{H}}_0$ diagonalization (and feasible for matrix dimensions in excess of $10^5$, with the sparsity preserved). Once the propagator is available, it becomes easy to evaluate the integral in the BRW master equation numerically. The resulting algorithm (essentially a numerical quadrature) is simple, general and computer-friendly, and has been implemented in the SPINACH library (MATLAB source code is available at http://spindynamics.org). Used together with the recently developed state space restriction tools for Liouville space [12-14], this puts the BRW relaxation theory treatment of large spin systems within reach.

## II. Rotational factorization of the dynamic Hamiltonian

To enable the treatment of arbitrary spin systems and to facilitate the evaluation of correlation functions in Section III, the Hamiltonian needs to be transformed into the irreducible spherical tensor form, which permits simple treatment of rotations [3,11,15-16] and gives a straightforward avenue to the rotational correlation functions [17-18]. For the traceless part of a bilinear interaction between spins $L$ and $S$:

$$\hat{T}_0^{(2)}(L,S) = +\sqrt{\frac{2}{3}}\left(\hat{L}_Z\hat{S}_Z - \frac{1}{4}\left(\hat{L}_+\hat{S}_- + \hat{L}_-\hat{S}_+\right)\right)$$

$$\hat{T}_{-1}^{(2)}(L,S) = +\frac{1}{2}\left(\hat{L}_Z\hat{S}_- + \hat{L}_-\hat{S}_Z\right), \quad \hat{T}_1^{(2)}(L,S) = -\frac{1}{2}\left(\hat{L}_Z\hat{S}_+ + \hat{L}_+\hat{S}_Z\right) \quad (3)$$

$$\hat{T}_{-2}^{(2)}(L,S) = +\frac{1}{2}\hat{L}_-\hat{S}_-, \quad \hat{T}_2^{(2)}(L,S) = +\frac{1}{2}\hat{L}_+\hat{S}_+$$

($\vec{\hat{L}}$ to be replaced with $\vec{B}$ in the case of Zeeman interaction and with $\vec{\hat{S}}$ in the case of quadratic interactions, such as quadrupolar or ZFS). For a traceless interaction tensor $\mathbf{A}$ with eigenvalues $\{a_{XX}, a_{YY}, a_{ZZ}\}$, written in its eigenframe:

$$\vec{\hat{L}}\cdot\mathbf{A}\cdot\vec{\hat{S}} = a_{XX}\hat{L}_X\hat{S}_X + a_{YY}\hat{L}_Y\hat{S}_Y + a_{ZZ}\hat{L}_Z\hat{S}_Z =$$
$$= \frac{2a_{ZZ} - (a_{XX} + a_{YY})}{\sqrt{6}}\hat{T}_0^{(2)}(L,S) + \frac{a_{XX} - a_{YY}}{2}\hat{T}_{-2}^{(2)}(L,S) + \frac{a_{XX} - a_{YY}}{2}\hat{T}_2^{(2)}(L,S) \quad (4)$$

Irreducible spherical tensors form the basis of irreducible representations of the rotation group [19] and therefore have very regular rotation properties [20]:

$$\hat{\hat{R}}\left(\hat{T}_k^{(l)}\right) = \sum_{m=-l}^{l} \hat{T}_m^{(l)} \mathfrak{D}_{mk}^{(l)}, \tag{5}$$

where $\hat{\hat{R}}$ denotes a rotation (which for our purposes is a superoperator) and $\mathfrak{D}_{m,k}^{(l)}$ are Wigner functions, accepting any rotation specification (Euler angles, quaternions *etc*) as an argument. For a multi-spin system in a rigid molecule, with the stochastic molecular rotation $\hat{\hat{R}}_{\text{mol}}(t)$ applied on top of static internal rotation $\hat{\hat{R}}_{\text{int}}$ for each interaction, we therefore have:

$$\hat{H} = \hat{H}_{\text{iso}} + \hat{\hat{R}}_{\text{mol}}(t) \begin{pmatrix} \sum_L \hat{\hat{R}}_{\text{int}}^L \left[ \frac{Rh_L}{2}\left(\hat{T}_2^{(2)}(B,L)+\hat{T}_{-2}^{(2)}(B,L)\right) + \frac{Ax_L}{\sqrt{6}}\hat{T}_0^{(2)}(B,L) \right] + \\ + \sum_{L,S} \hat{\hat{R}}_{\text{int}}^{LS} \left[ \frac{Rh_{LS}}{2}\left(\hat{T}_2^{(2)}(L,S)+\hat{T}_{-2}^{(2)}(L,S)\right) + \frac{Ax_{LS}}{\sqrt{6}}\hat{T}_0^{(2)}(L,S) \right] + \\ + \sum_S \hat{\hat{R}}_{\text{int}}^{SS} \left[ \frac{Rh_{SS}}{2}\left(\hat{T}_2^{(2)}(S,S)+\hat{T}_{-2}^{(2)}(S,S)\right) + \frac{Ax_{SS}}{\sqrt{6}}\hat{T}_0^{(2)}(S,S) \right] \end{pmatrix}, \tag{6}$$

where $\hat{H}_{\text{iso}}$ is the isotropic part of the Hamiltonian, axiality and rhombicity parameters for each interaction tensor are defined as:

$$Ax = 2a_{ZZ} - (a_{XX} + a_{YY}) \qquad Rh = a_{XX} - a_{YY}, \tag{7}$$

and the three terms in brackets in Equation (6) correspond to linear, bilinear and quadratic couplings within the spin system. After we apply the internal rotations using Equation (5), the Hamiltonian transforms into:

$$\hat{H} = \hat{H}_{\text{iso}} + \hat{\hat{R}}_{\text{mol}}(t) \sum_L \sum_{m=-2}^{2} \hat{T}_m^{(2)}(B,L) \Phi_m(B,L) + \hat{\hat{R}}_{\text{mol}}(t) \sum_{LS} \sum_{m=-2}^{2} \hat{T}_m^{(2)}(L,S) \Phi_m(L,S) + \\ + \hat{\hat{R}}_{\text{mol}}(t) \sum_S \sum_{m=-2}^{2} \hat{T}_m^{(2)}(S,S) \Phi_m(S,S), \tag{8}$$

where the interaction amplitudes and orientations in the molecular frame have been collected into internal orientation coefficients:

$$\Phi_m(B,L) = \frac{Rh_L}{2}\left(\mathfrak{D}_{m,-2}^{(2)}(B,L) + \mathfrak{D}_{m,2}^{(2)}(B,L)\right) + \frac{Ax_L}{\sqrt{6}}\mathfrak{D}_{m,0}^{(2)}(B,L)$$

$$\Phi_m(L,S) = \frac{Rh_{LS}}{2}\left(\mathfrak{D}_{m,-2}^{(2)}(L,S) + \mathfrak{D}_{m,2}^{(2)}(L,S)\right) + \frac{Ax_{LS}}{\sqrt{6}}\mathfrak{D}_{m,0}^{(2)}(L,S) \tag{9}$$

$$\Phi_m(S,S) = \frac{Rh_{SS}}{2}\left(\mathfrak{D}_{m,-2}^{(2)}(S,S) + \mathfrak{D}_{m,2}^{(2)}(S,S)\right) + \frac{Ax_{SS}}{\sqrt{6}}\mathfrak{D}_{m,0}^{(2)}(S,S),$$

where $\mathfrak{D}_{m,k}^{(2)}$ are time-independent Wigner functions [20] specifying the orientation of the corresponding interactions in the molecular frame. Finally, after we apply Equa-tion (5) the overall molecular rotation (which is assumed to be time-dependent) in Equation (8), we obtain

$$\hat{H} = \hat{H}_{iso} + \sum_{m=-2}^{2}\sum_{k=-2}^{2} \mathfrak{D}_{km}^{(2)}(t)\hat{Q}_{km}, \qquad (10)$$

where all information about the amplitudes and internal orientations of all interactions has been packaged into 25 static operators:

$$\hat{Q}_{km} = \sum_{L}\Phi_m(B,L)\hat{T}_k^{(2)}(B,L) + \sum_{LS}\Phi_m(L,S)\hat{T}_k^{(2)}(L,S) + \sum_{S}\Phi_m(S,S)\hat{T}_k^{(2)}(S,S). \qquad (11)$$

We will call these operators *rotational basis*, they are the generalization of the classical 'dipolar alphabet' [1] to the case of multiple arbitrarily oriented rhombic interactions. Importantly, the software implementations of Equation (10) are relatively easy to test and debug, because, unlike relaxation theory, easily verifiable standard results are available for spin system rotations.

## III. BRW theory in Liouville space

This section gives the Liouville space version of the well known [1-2,4-6] route towards the BRW relaxation superoperator, taking advantage of the general rotational factorization given by Equation (10). We start with the adjoint representation of the Liouville - von Neumann equation:

$$\frac{\partial \hat{\rho}(t)}{\partial t} = -i\left(\hat{\hat{H}}_0 + \hat{\hat{H}}_1(t)\right)\hat{\rho}(t), \qquad \hat{\hat{H}} = \mathrm{ad}\hat{H} = [\hat{H},\cdot] = \hat{H}\otimes\hat{E} - E\otimes\hat{H}^{\mathrm{T}}$$

$$\exp\left(-i\hat{\hat{H}}\Delta t\right)\hat{\rho} = \left[\sum_{n=0}^{\infty}\frac{(-i\Delta t)^n}{n!}\hat{\hat{H}}^n\right]\hat{\rho} = \sum_{n=0}^{\infty}\frac{(-i\Delta t)^n}{n!}[\hat{H},[\hat{H},[...[\hat{H},\hat{\rho}]]]] \qquad (12)$$

with the assumption (without loss of generality) that the ensemble average of the dynamic part $\hat{\hat{H}}_1(t)$ is zero. The transformation to the interaction representation (denoted with $R$ superscript on the operators) with respect to the static Hamiltonian is then performed using the following relations:

$$\hat{\sigma}(t) = e^{i\hat{\hat{H}}_0 t}\hat{\rho}(t) \qquad \hat{\hat{H}}_1^{\mathrm{R}}(t) = e^{i\hat{\hat{H}}_0 t}\hat{\hat{H}}_1(t)e^{-i\hat{\hat{H}}_0 t} \qquad (13)$$

with the result that the static part formally vanishes from Equation (12) and is replaced by the unitary transformation prescribed by Equation (13):

$$\frac{\partial \hat{\rho}(t)}{\partial t} = -i\left(\hat{\hat{H}}_0 + \hat{\hat{H}}_1(t)\right)\hat{\rho}(t) \quad \Rightarrow \quad \frac{\partial \hat{\sigma}(t)}{\partial t} = -i\hat{\hat{H}}_1^{\mathrm{R}}(t)\hat{\sigma}(t), \qquad (14)$$

this can be verified by direct inspection. We can formally integrate this equation

$$\hat{\sigma}(t) = \hat{\sigma}(0) - i\int_0^t \hat{\hat{H}}_1^R(t')\hat{\sigma}(t')dt' \quad (15)$$

and re-substitute the result back into Equation (14) to yield:

$$\frac{\partial \hat{\sigma}(t)}{\partial t} = -i\hat{\hat{H}}_1^R(t)\hat{\sigma}(0) - \int_0^t \hat{\hat{H}}_1^R(t)\hat{\hat{H}}_1^R(t')\hat{\sigma}(t')dt' \quad (16)$$

If ensemble averaging is performed on this equation, the first term on the right hand side vanishes because

$$\left\langle \hat{\hat{H}}_1^R(t)\hat{\sigma}(0) \right\rangle = e^{i\hat{\hat{H}}_0 t}\left\langle \hat{\hat{H}}_1(t) \right\rangle e^{-i\hat{\hat{H}}_0 t}\hat{\sigma}(0) = 0 \quad (17)$$

(the angle brackets denote the ensemble averaging), and we are left with

$$\frac{\partial \langle \hat{\sigma}(t) \rangle}{\partial t} = -\int_0^t \left\langle \hat{\hat{H}}_1^R(t)\hat{\hat{H}}_1^R(t')\hat{\sigma}(t') \right\rangle dt' \quad (18)$$

We can now use the rotational factorization derived in Section II and take advantage of the fact that it hides the details of the interactions within the spin system, leaving only the overall molecular rotation explicit:

$$\hat{\hat{H}}_1(t) = \sum_{km} \mathfrak{D}_{km}^{(2)}(t)\hat{\hat{Q}}_{km} = \sum_{km} \mathfrak{D}_{km}^{(2)*}(t)\hat{\hat{Q}}_{km}^\dagger$$

$$\frac{\partial \langle \hat{\sigma}(t) \rangle}{\partial t} = -\sum_{kmpq} \int_0^t \left\langle \mathfrak{D}_{km}^{(2)}(t)\mathfrak{D}_{pq}^{(2)*}(t')\hat{\hat{Q}}_{km}^R(t)\hat{\hat{Q}}_{pq}^{R\dagger}(t')\sigma(t') \right\rangle dt' \quad (19)$$

(to facilitate subsequent treatment, one copy of the hermitian $\hat{\hat{H}}_1^R(t)$ superoperator has been pasted in a hermitian conjugate form). We need to introduce several significant assumptions at this point [4,21]. Firstly, we shall assume that the spin system dynamics is uncorrelated with the noise that is driving $\hat{\hat{H}}_1^R(t)$ – it would allow us to take ensemble averages separately for the Liouvillian and the state vector part in Equation (19):

$$\left\langle \mathfrak{D}_{km}^{(2)}(t)\mathfrak{D}_{pq}^{(2)*}(t')\hat{\hat{Q}}_{km}^R(t)\hat{\hat{Q}}_{pq}^{R\dagger}(t')\sigma(t') \right\rangle = $$
$$= \left\langle \mathfrak{D}_{km}^{(2)}(t)\mathfrak{D}_{pq}^{(2)*}(t') \right\rangle \hat{\hat{Q}}_{km}^R(t)\hat{\hat{Q}}_{pq}^{R\dagger}(t')\langle \sigma(t') \rangle \quad (20)$$

(the interaction representations of the rotational basis operators evolve deterministically and do not change upon ensemble averaging). It is convenient to introduce a separate symbol for the rotational correlation functions, which, in the general anisotropic tumbling case can be different for different values of the *k,m,p,q* indices [21]:

$$G_{kmpq}(t,t') = \left\langle \mathfrak{D}_{km}^{(2)}(t) \mathfrak{D}_{pq}^{(2)*}(t') \right\rangle \tag{21}$$

The second significant assumption is that the noise in the system is stationary, and therefore the correlation functions only depend on the time separation between the Wigner functions to be correlated, therefore:

$$G_{kmpq}(t-t') = G_{kmpq}(t'-t) = \left\langle \mathfrak{D}_{km}^{(2)}(t) \mathfrak{D}_{pq}^{(2)*}(t') \right\rangle \tag{22}$$

The third assumption is that these correlation functions decay so fast on the time scale of the overall spin system evolution, that the latter barely occurs and it is permissible to take the state vector out of the integral:

$$\frac{\partial \langle \hat{\sigma}(t) \rangle}{\partial t} = -\left[ \sum_{kmpq} \int_0^t G_{kmpq}(t-t') \hat{\hat{Q}}_{km}^{\mathrm{R}}(t) \hat{\hat{Q}}_{pq}^{\mathrm{R}\dagger}(t') dt' \right] \langle \hat{\sigma}(t) \rangle \tag{23}$$

(because the implicit coarse-grained propagation schemes are known to be more stable than the explicit ones [22], we take $\hat{\sigma}(t)$ out of the integral, rather than $\hat{\sigma}(0)$). We shall also drop the angle brackets on the density matrix for convenience. Using Equation (13) to return to the Schrödinger representation, we get:

$$\frac{\partial \hat{\rho}(t)}{\partial t} = -i\hat{\hat{H}}_0 \hat{\rho}(t) - \left[ \sum_{kmpq} \int_0^t G_{kmpq}(t-t') \hat{\hat{Q}}_{km} e^{-i\hat{H}_0(t-t')} \hat{\hat{Q}}_{pq}^\dagger e^{i\hat{H}_0(t-t')} dt' \right] \hat{\rho}(t) \tag{24}$$

the integrand in this equation only depends on the time difference between *t* and *t'*, and a variable substitution $\tau = t - t'$ results in

$$\frac{\partial \hat{\rho}(t)}{\partial t} = -i\hat{\hat{H}}_0 \hat{\rho}(t) - \left[ \sum_{kmpq} \int_0^t G_{kmpq}(\tau) \hat{\hat{Q}}_{km} e^{-i\hat{H}_0 \tau} \hat{\hat{Q}}_{pq}^\dagger e^{i\hat{H}_0 \tau} d\tau \right] \hat{\rho}(t) \tag{25}$$

Finally, we note that, because $G_{kmpq}(\tau)$ functions decay very fast within the $[0,t]$ time interval, we can extend the upper integration limit to infinity. The relaxation superoperator can then be identified as

$$\hat{\hat{R}} = -\sum_{kmpq} \int_0^\infty G_{kmpq}(\tau) \hat{\hat{Q}}_{km} e^{-i\hat{H}_0 \tau} \hat{\hat{Q}}_{pq}^\dagger e^{i\hat{H}_0 \tau} d\tau \tag{26}$$

and relaxation to a user-specified thermal equilibrium may be set up, if necessary, as a one-way coupling to the unit operator as described by Levitt and Di Bari [23].

## IV. Efficient evaluation of the time integral

We are faced with the problem of computing superoperator-valued integrals of the following general type:

$$\int_0^\infty G(\tau) e^{-i\hat{\hat{H}}_0 \tau} \hat{\hat{Q}} e^{i\hat{\hat{H}}_0 \tau} d\tau, \qquad (27)$$

where $G(\tau)$ is a correlation function and matrices $\hat{\hat{H}}_0$ and $\hat{\hat{Q}}$ are very sparse. Historically, the standard way of evaluating this integral was to diagonalize $\hat{\hat{H}}_0$ and expand $\hat{\hat{Q}}$ in its eigenstates, at which point it splits into a collection of analytical Fourier transforms [1-2,4]. This is easy for small spin systems, but with matrix dimensions in excess of $10^5$ in large-scale simulations [12-14], this approach becomes impractical, the greatest constraint being memory – eigenvectors of sparse matrices are in most cases dense. A more detailed analysis is therefore in order on what could be done to avoid $\hat{\hat{H}}_0$ diagonalization.

Firstly, we shall consider the restraints put on the behaviour of the various parts of Equation (27) by the validity conditions of BRW theory. The theory is rooted in the generalized cumulant expansion for the effective step Liouvillian [10]

$$\ln \left\langle \exp\left( -i \int_0^{\Delta t} \hat{\hat{H}}_1^R(t) dt \right) \right\rangle = \sum_{n=1}^\infty (-i)^n \int_0^{\Delta t} dt_1 \int_0^{t_1} dt_2 \ldots \int_0^{t_{n-1}} dt_n \left\langle \hat{\hat{H}}_1^R(t_1) \hat{\hat{H}}_1^R(t_2) \ldots \hat{\hat{H}}_1^R(t_n) \right\rangle_C, \quad (28)$$

where convergence is guaranteed if the 2-norm (defined as the largest singular value [24]) of the dynamic Hamiltonian satisfies:

$$\left\| \hat{\hat{H}}_1^R(t) \right\| \Delta t = \left\| \hat{\hat{H}}_1(t) \right\| \Delta t < 1. \qquad (29)$$

This condition may be derived from the root test [25] on the convergence of the series in Equation (28):

$$\limsup_{n \to \infty} \sqrt[n]{\left\| (-i)^n \int_0^{\Delta t} dt_1 \int_0^{t_1} dt_2 \ldots \int_0^{t_{n-1}} dt_n \left\langle \hat{\hat{H}}_1^R(t_1) \hat{\hat{H}}_1^R(t_2) \ldots \hat{\hat{H}}_1^R(t_n) \right\rangle_C \right\|} \leq$$

$$\leq \limsup_{n \to \infty} \sqrt[n]{\int_0^{\Delta t} dt_1 \int_0^{t_1} dt_2 \ldots \int_0^{t_{n-1}} dt_n \left\| \left\langle \hat{\hat{H}}_1^R(t_1) \hat{\hat{H}}_1^R(t_2) \ldots \hat{\hat{H}}_1^R(t_n) \right\rangle_C \right\|} \leq \qquad (30)$$

$$\leq \limsup_{n \to \infty} \sqrt[n]{\Delta t^n \left\| \hat{\hat{H}}_1^R(t) \right\|^n} = \left\| \hat{\hat{H}}_1^R(t) \right\| \Delta t = \left\| \hat{\hat{H}}_1(t) \right\| \Delta t < 1$$

where the norm $\|\hat{\hat{H}}_1(t)\|$ is defined as the largest norm that $\hat{\hat{H}}_1(t)$ has within the $[0, \Delta t]$ interval. Equation (29) is quite pessimistic – in practice, the alternating signs in Equation (28) are likely to improve convergence, but for our purposes it simply means a safer accuracy estimate. BRW theory truncates the cumulant expansion at the second term, so we must additionally have:

$$\left\| \int_0^{\Delta t} dt_1 \int_0^{t_1} dt_2 \int_0^{t_2} dt_3 \left\langle \hat{\hat{H}}_1^R(t_1) \hat{\hat{H}}_1^R(t_2) \hat{\hat{H}}_1^R(t_3) \right\rangle \right\| \ll \left\| \int_0^{\Delta t} dt_1 \int_0^{t_1} dt_2 \left\langle \hat{\hat{H}}_1^R(t_1) \hat{\hat{H}}_1^R(t_2) \right\rangle \right\|, \quad (31)$$

in practice by at least two orders of magnitude, if the intention is to match the best available experimental accuracy, which is about 1%. Equation (31) uses the fact that, for centred stochastic processes, the second and third cumulants are equal to the second and third moments [9-10]. With this in place, Equation (29) becomes significantly more stringent:

$$\left\|\hat{\hat{H}}_1(t)\right\| \Delta t < 10^{-2}. \quad (32)$$

The other approximation that was made in Section III – taking the integral limit to infinity – must also have a negligible effect on the accuracy, meaning that

$$\left\| \int_{\Delta t}^{\infty} \left\langle \hat{\hat{H}}_1^R(0) \hat{\hat{H}}_1^R(t) \right\rangle dt \right\| \ll \left\| \int_0^{\infty} \left\langle \hat{\hat{H}}_1^R(0) \hat{\hat{H}}_1^R(t) \right\rangle dt \right\| \quad (33)$$

by the same two orders of magnitude (greater accuracy can, of course, be obtained by increasing this cut-off tolerance). For an exponentially decaying correlation functions behaving asymptotically as $\exp(-t/\tau_C)$, this yields $\tau_C < \Delta t / 5$. Together with Equation (29) this produces the condition under which BRW theory is accurate to 1%:

$$\left\|\hat{\hat{H}}_1(t)\right\| \tau_C^{\max} < 2 \cdot 10^{-3}, \quad (34)$$

where $\tau_C^{\max}$ is the longest correlation time in the system. This is the point (about $\tau_C^{\max} \sim 20$ ns in common protein dipolar networks and $\tau_C^{\max} \sim 20$ ps in common aromatic radicals, actual numbers varying greatly from system to system) beyond which BRW theory is likely to break.

This has implications for the integral in Equation (27) – if the theory is not used outside of its validity range, we do not have to propagate $e^{-i\hat{\hat{H}}_0 \tau} \hat{\hat{Q}} e^{i\hat{\hat{H}}_0 \tau}$ very far – even with the Hamiltonian superoperator as large as the full NMR $\hat{\hat{H}}_0$, a few hundred nanoseconds is very manageable, the primary reason being that the step propagator $\exp[-i\hat{\hat{H}}_0 \Delta\tau]$ is cheap to compute when $\|\hat{\hat{H}}_0\| \Delta\tau \leq 1$, because the various approximations to $\exp[-i\hat{\hat{H}}_0 \Delta\tau]$ converge very rapidly [26-27]:

$$e^{-i\hat{\hat{H}}_0 \Delta\tau} = \frac{\sum_{j=0}^{p} \frac{(p+q-j)!p!}{(p+q)!j!(p-j)!}\left(-i\hat{\hat{H}}_0 \Delta\tau\right)^j}{\sum_{j=0}^{q} \frac{(p+q-j)!p!}{(p+q)!j!(p-j)!}\left(i\hat{\hat{H}}_0 \Delta\tau\right)^j} + O\left(\left\|i\hat{\hat{H}}_0 \Delta\tau\right\|^{p+q+1}\right), \quad (35)$$

$$e^{-i\hat{\hat{H}}_0 \Delta\tau} = \sum_{n=0}^{p} (2-\delta_{n0}) i^n J_n(1) T_n\left(-\hat{\hat{H}}_0 \Delta\tau\right) + O\left(\left\|i\hat{\hat{H}}_0 \Delta\tau\right\|^{p+1}\right), \quad (36)$$

$$e^{-i\hat{\hat{H}}_0 \Delta\tau} = \sum_{n=0}^{p} \frac{(-i\hat{\hat{H}}_0 \Delta\tau)^n}{n!} + O\left(\left\|i\hat{\hat{H}}_0 \Delta\tau\right\|^{p+1}\right). \quad (37)$$

where $\delta_{nk}$ is the Kronecker delta, $J_n(x)$ are Bessel functions and $T_n(x)$ are Chebyshev polynomials. The Chebyshev approximation in Equation (36) and Taylor approximation in Equation (37) are generally faster for large sparse matrices, because they avoid computing a matrix inverse, which is required for the Padé approximation in Equation (35). Importantly, low powers of sparse matrices are also sparse, and sparsity can be improved further by dropping the insignificant elements from the non-zero index after each multiplication operation. Table 1 gives some examples for common NMR systems and Figure 1 gives a timing comparison for a series of large Liouvillian matrices.

If numerical step propagators are cheap and the number of steps required is small, the obvious way to evaluate the integral in Equation (27) is by a numerical quadrature. For efficiency reasons, this cannot be a variable-step method, such as Gauss-Legendre quadrature – we would ideally like to only compute the time step propagator $e^{-i\hat{\hat{H}}_0 \Delta t}$ once and then reuse it. For a fixed-step method, a reasonable trade-off between expense and accuracy is offered by Boole's $O(h^7)$ quadrature [28]:

$$\int_{t_1}^{t_5} f(t) dt = \frac{2h}{45}\left(7f(t_1) + 32f(t_2) + 12f(t_3) + 32f(t_4) + 7f(t_5)\right) - \frac{8}{945} h^7 f^{(6)}(t_1 + \theta(t_5 - t_1)), \quad (38)$$

where $h$ is the point spacing, $t_k = t_{k-1} + h$ and $\theta$ is a number between 0 and 1. The accuracy of integration is determined by the cut-off time (since the integral in Equation (27) is indefinite) and the point spacing $h$. If $\xi$ is the desired relative accuracy and $\tau_C$ is the characteristic decay time of $G(t)$, then the upper integration limit $t_{max}$ should be at least $\tau_C \ln(1/\xi)$ and the number of integration steps (a conservative estimate) should be

$$n_{steps} > \frac{1}{10\sqrt[6]{\xi}} \left( \frac{\tau_C \ln(1/\xi)}{\min\{\tau_C, \|H_0\|^{-1}\}} \right)^{7/6}. \quad (39)$$

In practical simulations $\sim 10^{-4}$ relative accuracy is usually sufficient, yielding $t_{max} = 9\tau_C$ and $n_{steps} = 8$. This puts the total cost of computing the full relaxation superoperator to about 100 sparse matrix multiplications in Liouville space.

## V. Illustrations

Table 1 illustrates the considerable difference between the time it takes to compute a step propagator and the time required to diagonalize the Hamiltonian superoperator. The primary advantage of exponentiation comes from the fact that low powers of sparse matrices in Equation (37) are also sparse. Because the operator density drops off rapidly with the size of the spin system [29] (this is also true in restricted state spaces [13-14]) the sparse multiplications are fast. In contrast, the eigenvector array generated in the diagonalization is dense. Another advantage comes from the small memory footprint of sparse matrices *versus* the need to store the full eigenvector array in the case of diagonalization. For this reason, the entire code base of the SPINACH library (including modules other than the rotations and BRW theory that this paper deals with) does not contain a single diagonalization operation.

With the full relaxation superoperator in place, interesting relaxation-driven experiments can be simulated accurately for large spin systems in liquid state using the state space restriction techniques that we had previously reported [13-14]. An example of a NOESY spectrum of the 22-spin system of strychnine is shown in Figure 2 (isotropic tumbling, $\tau_C = 200$ ps, mixing time set to 500 ms). A COSY spectrum is also shown to illustrate the fact that all scalar couplings are handled accurately. The relaxation superoperator used to compute both spectra includes all chemical shielding anisotropies, all dipolar couplings between spins closer than 4 Angstroms and all cross-correlations between the orientations of all interaction tensors – as a full relaxation superoperator rightly should.

The accuracy is further illustrated in Figure 3 using the very well characterised cross-correlation behaviour in the $^{19}$F relaxation of 1-fluoro-2,4-dinitrobenzene (FDNB). The agreement with experiment is almost perfect (*c.f.* Figure 5 in the experimental paper by Grace and Kumar [30]) – a tribute to the accuracy of modern DFT methods as well as BRW theory implementation described above. The theoretical $\langle \hat{F}_Z \rangle$ relaxation rate is 0.384 s$^{-1}$ and the DD-CSA cross-correlation rate between $\langle \hat{F}_Z \rangle$ and $\langle 2\hat{H}_Z^{(2)}\hat{F}_Z \rangle$ is –0.068 s$^{-1}$ *versus* 0.4 s$^{-1}$ and –0.067 s$^{-1}$ determined experimentally [30]. The minor difference is likely due to the slight anisotropy in the rotational diffusion tensor of FDNB.

An ESR spectroscopy example (*para*-fluorotoluene radical, isotropic tumbling, $\tau_C = 50$ ps) is given in Figure 4. The transverse relaxation in this system features a significant contribution from the cross-correlation between the anisotropies of hyperfine and *g*-tensors – this effect is clearly visible in the simulated spectrum.

## VI. Conclusions

It appears that, in the context of spin relaxation theory, matrix exponentials giving small step time propagators are significantly cheaper computationally than matrix diagonalization. This suggests an alternative path (using numerical integration rather than Fourier transforms) through Bloch-Redfield-Wangsness theory, which is presented above and shown to be much faster, particularly for large spin systems.

## Acknowledgements

The author would like to thank Hannah Hogben and Peter Hore for stimulating discussions. The project is funded by the EPSRC (EP/F065205/1, EP/H003789/1) and supported by the Oxford e-Research Centre.

# Figure captions

**Figure 1**    Wall clock time (contemporary uniprocessor workstation) taken for the calculation of $\exp(-iL\Delta t)$ with $\Delta t \approx (1/2)\|\hat{\hat{L}}\|^{-1}$ for a series of linear spin chains containing between 5 and 100 proton spins with strong nearest-neighbour *J*-coupling (with state space restriction [13-14] up to, and including, four-spin states between directly coupled spins). Of the two polynomial approximations, the Chebyshev method, even though it requires fewer iterations, is slower on the wall clock than Taylor series due to greater memory requirements.

**Figure 2**    Theoretical NOESY and DQF-COSY spectra of strychnine (22 proton spins, assuming a rigid molecular structure at the DFT energy minimum) calculated using explicit time propagation in restricted Liouville space with the relaxation superoperator computed as described in the main text. $^1$H chemical shielding tensors, distances and *J*-couplings were obtained from a GIAO DFT B3LYP/EPR-II calculation.

**Figure 3**    Theoretical 376 MHz $^{19}$F inversion-recovery NMR spectra of 1-fluoro-2,4-dinitrobenzene plotted as a function of mixing time. $^{19}$F and $^1$H chemical shielding tensors and couplings were obtained from a separate GIAO DFT B3LYP/EPR-II calculation.

**Figure 4**    Theoretical X-band ESR spectrum of *para*-fluorotoluene radical, computed using explicit time propagation in Liouville space with the relaxation superoperator computed as described in the main text. Anisotropies of all interaction tensors were obtained from a GIAO DFT B3LYP/EPR-II calculation. The line width pattern (increasing from left to right) is typical for $\Delta g$-$\Delta a$ cross-correlated relaxation.

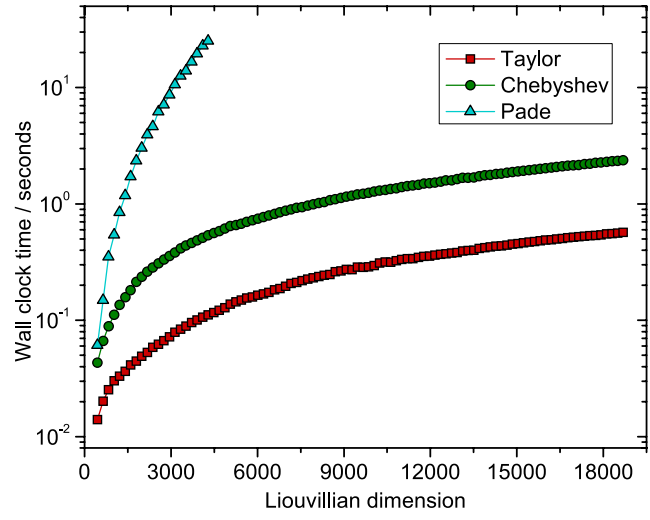

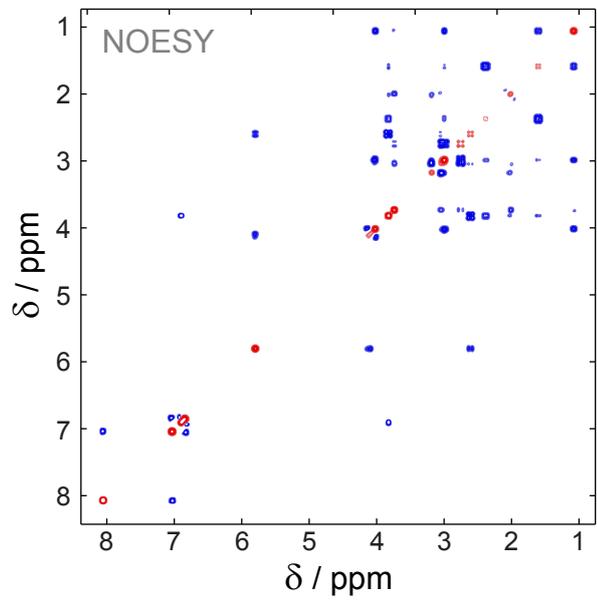 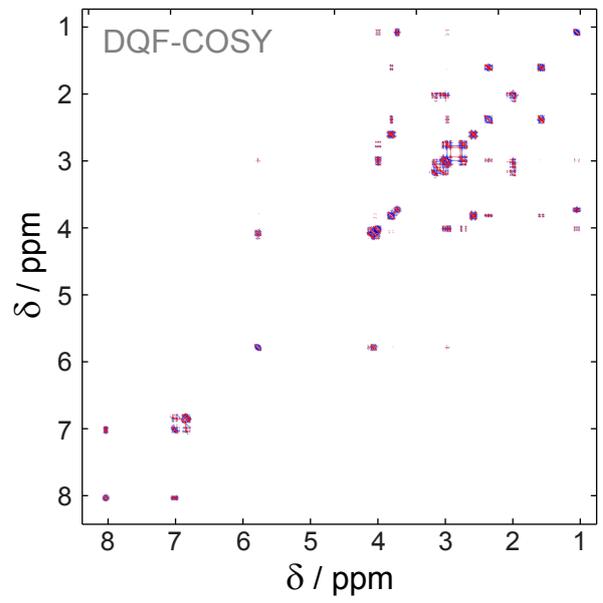

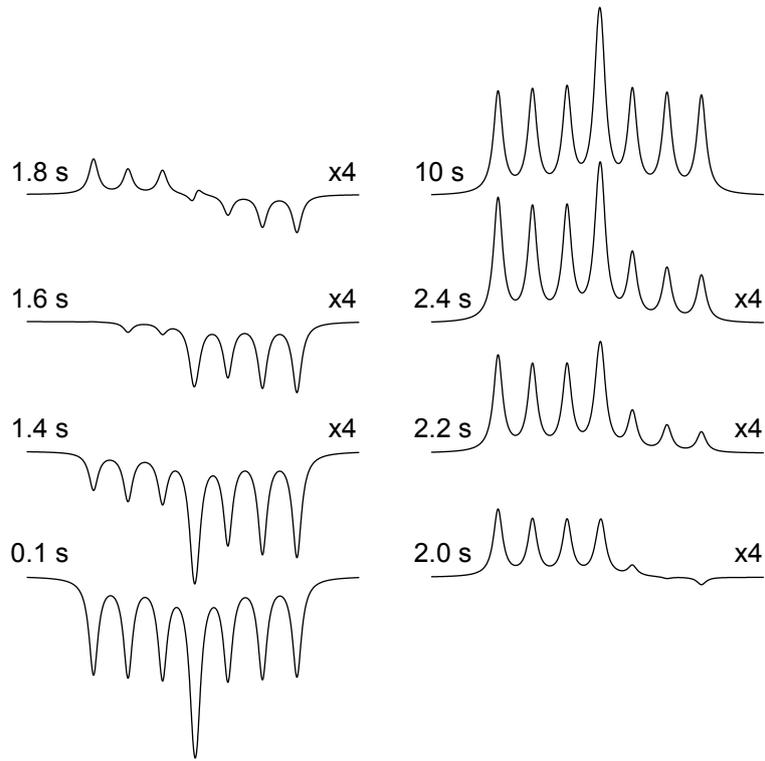

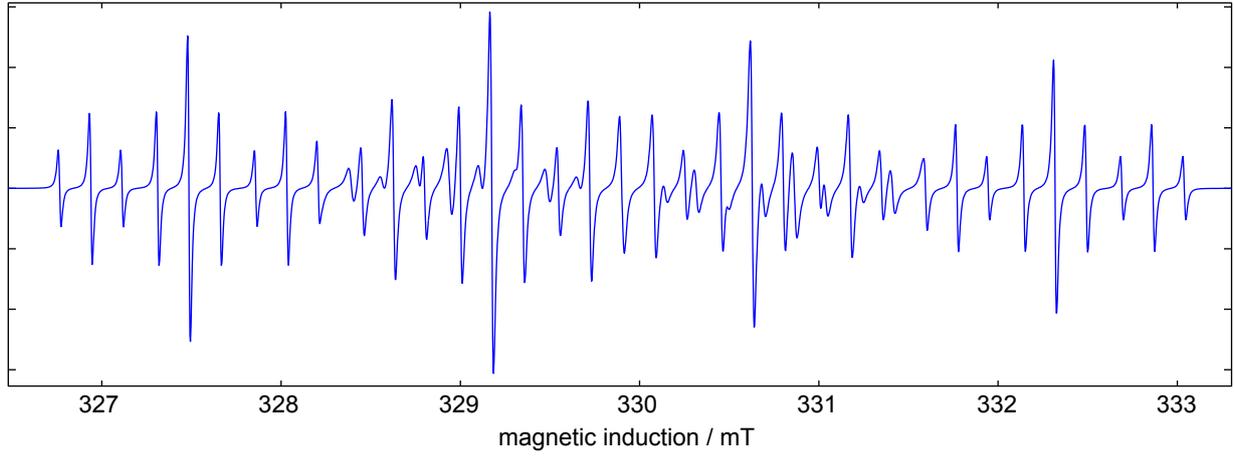

**Table 1.** CPU time statistics for diagonalization and exponentiation of static Hamiltonian commutation superoperators commonly encountered in liquid-state NMR spectroscopy.

| Spin system | State space | State space dimension | Wall clock time, $\hat{\hat{H}}_0$ diagonalization | Wall clock time, $\exp(-i\hat{\hat{H}}_0 \Delta t)$ [d] |
|---|---|---|---|---|
| FNDB ($^1$H,$^{19}$F) | complete | 256 | 0.06 s | 0.01 s |
| glycine ($^1$H,$^{13}$C) | complete ($A_{1g}$ of $S_3 \times S_2$)[a] | 3200 | 120 s | 0.4 s |
| isoleucine ($^1$H) | SSR-5,3[b] ($A_{1g}$ of $S_3 \times S_3$)[c] | 15357 | – | 2.8 s |
| strychnine ($^1$H) | SSR-5,3 | 32818 | – | 5.4 s |
| sucrose ($^1$H,$^{13}$C) | SSR-5,3 | 88393 | – | 15.1 s |

[a] Fully symmetric irreducible representation of the $S_3 \times S_2$ symmetry group (three equivalent protons at nitrogen and two equivalent protons at $C_\alpha$).

[b] State space restriction up to (and including) five-spin orders between directly *J*-coupled spins and three-spin orders between all spins within 4 Angstroms of each other.

[c] Fully symmetric irreducible representation of the $S_3 \times S_3$ symmetry group (two groups of three equivalent protons at the two methyl carbons).

[d] Taylor approximation. The time step is chosen to satisfy $\Delta t = \| \hat{\hat{H}}_0 \|^{-1}$ with *Matlab*'s `normest` sparse norm estimator used to compute $\| \hat{\hat{H}}_0 \|$.